\documentclass[aps,showpacs,prd,twocolumn,nofootinbib,nobibnotes,]{revtex4-2}

\usepackage{graphicx}
\usepackage{amssymb}
\usepackage{amsmath}
\usepackage{color}
\usepackage{float}
\usepackage{ulem}
\usepackage{accents}
\usepackage{graphicx}
\usepackage{graphicx}
\usepackage{amsfonts}
\usepackage[colorlinks=true,
pdfstartview=FitV,linkcolor=blue,
citecolor=blue,urlcolor=blue,breaklinks=true]
{hyperref}
\usepackage{array}
\usepackage{float}
\usepackage{placeins}
\usepackage[dvipsnames]{xcolor}
\usepackage{csquotes}
\usepackage{bbold}
\usepackage{units}
\usepackage{makecell}
\usepackage{enumitem}
\usepackage{dsfont}
\usepackage{upgreek}

\newcolumntype{C}[1]{>{\centering\arraybackslash}m{#1}}

\renewcommand{\eqref}[1]{\mbox{Eq.~(\ref{#1})}}

\definecolor{ForestGreen}{rgb}{0.13,0.55,0.13}

\usepackage{fixmath}

\newcommand{\orcid}[1]{\href{https://orcid.org/#1}{\includegraphics[width=10pt]{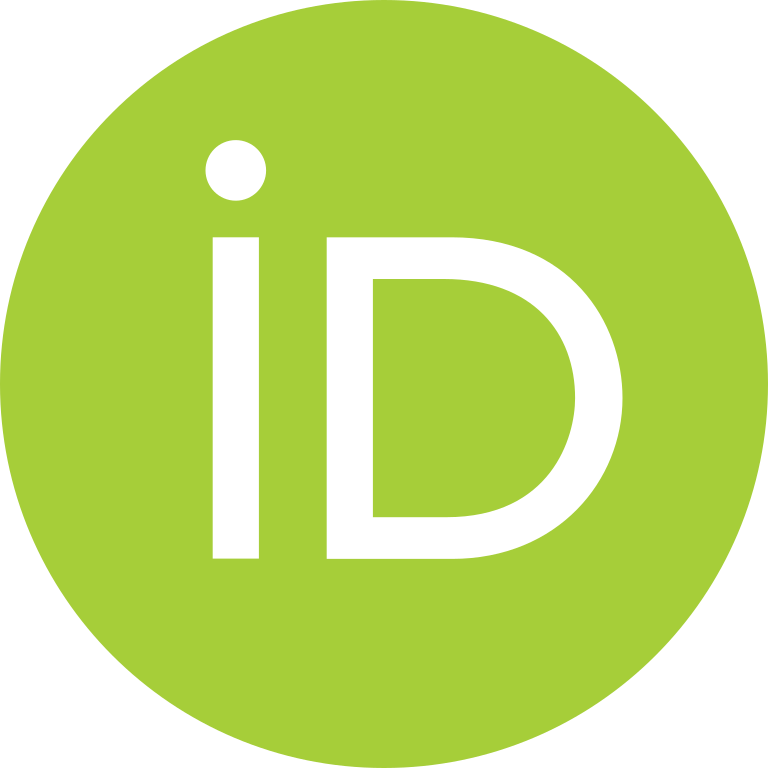}}}

\begin{document}
	
	\title{Optical effects in unmagnetized cold plasmas by a chiral axion factor}

\author{Filipe S. Ribeiro\orcid{0000-0003-4142-4304}$^a$}
\email{filipe.ribeiro@discente.ufma.br, filipe99ribeiro@hotmail.com}
	\author{Pedro D. S. Silva\orcid{0000-0001-6215-8186}$^a$}
	\email{pedro.dss@ufma.br, pdiego.10@hotmail.com}
	\author{Manoel M. Ferreira Jr.\orcid{0000-0002-4691-8090}$^{a,b}$}
	\email{\textcolor{black}{Corresponding author:} manojr.ufma@gmail.com, \\ manoel.messias@ufma.br}
		\affiliation{$^a$Programa de P\'{o}s-graduaç\~{a}o em F\'{i}sica, Universidade Federal do Maranh\~{a}o, Campus
		Universit\'{a}rio do Bacanga, S\~{a}o Lu\'is, {Maranhão} 65080-805, Brazil}
	\affiliation{$^b$Departamento de F\'{i}sica, Universidade Federal do Maranh\~{a}o, Campus
		Universit\'{a}rio do Bacanga, S\~{a}o Lu\'is, {Maranhão} 65080-805, Brazil}

\begin{abstract}
	Unmagnetized cold plasma modes are investigated in the context of the chiral Maxwell-Carroll-Field-Jackiw (MCJF) electrodynamics, where the axion chiral factor acts retrieving some typical properties of magnetized plasmas. The Maxwell equations are rewritten for a cold, uniform, and collisionless fluid plasma model, allowing us to determine the dispersion relation, new refractive indices, and propagating modes. We find four distinct refractive indices modified by the purely timelike CFJ background that plays the magnetic conductivity chiral parameter role associated with right-circularly polarized [RCP] and left-circularly polarized [LCP] waves. For each refractive index, the propagation and absorption zones are determined and illustrated for some specific parameter values. Modified RCP and LCP helicons are found in the low-frequency regime. The optical behavior is investigated, revealing that the chiral factor induces birefringence, measured in terms of the rotatory power (RP). The dichroism coefficient is carried out for the absorbing zones. The negative refraction zones may enhance the involved rotatory power, yielding RP sign reversion, a feature of rotating plasmas and MCFJ chiral plasmas. Charge density oscillations and Langmuir waves are also discussed, revealing no modified dispersion relation due to the chiral axion factor.

\end{abstract}
\maketitle

\section{Introduction \label{themodel1}}

The properties of electromagnetic modes in a cold magnetized plasma are described by the standard Maxwell equations written in continuous media \cite{STURROK, Bittencourt,chapter-8,Boyd},  which provides the basis for the radio wave propagation in the ionosphere, where the earth magnetic field plays a key role \cite{Appleton32,Appleton32B,Hartree,RATCLIFF2}. Although most plasma investigations occur in the presence of an averaged magnetic field, non-magnetized plasmas have also been explored in some extension, including topics such as vortex beams propagation \cite{HLi,HLi2, JWang, LWu}, dusty plasma systems \cite{Vladimirov, Vladimirov2, Varma, Shukla, MRAmin, Dodin},  nonlinear propagating waves \cite{Decker, Vranjes, Mahmood2, Ghosh, Chakrabarti} and self-focusing beams \cite{Walia, Walia2}. In the astrophysical context, electrodynamics of unmagnetized plasma was investigated in an expanding universe scenario \cite{Dettmann}. Recently, unmagnetized plasmas have been considered around rotating black holes to study the properties of shadows \cite{Koide, Perlick,Badía, Briozzo,Kumar}.

Cold plasma theory is usually described in terms of small fluctuations in an equilibrium configuration by using the Fourier formalism. When one considers that the ions are infinitely massive and  thermal/collisional effects can be neglected, the cold plasma scenario \cite{Boyd} is well addressed by a linearized system of equations for the relevant physical  fluctuations around average values (constant in time and space). The corresponding diagonal dielectric tensor is found,
\begin{equation}
		\varepsilon_{ij}  (\omega)=\varepsilon_{0} \left(1-\omega_{p}^{2}/ \omega^{2} \right)	\delta_{ij},
		\label{Epip1}
	\end{equation}
where $\omega_{p}^2=n_{0}q^{2}/{m\epsilon_{0}}$ is the plasma frequency.

The optical activity, manifested as the rotation of the polarization vector of the propagating wave (birefringence), is one optical signature of magnetized plasmas where the external field implies anisotropy, inducing the so-called Faraday effect \cite{Porter}. Optical activity has been an important topic in condensed matter chiral systems \cite{TangPRL}, being a feature of bi-isotropic \cite{Sihvola2, Ougier} and bi-anisotropic electrodynamics \cite{Itin,Aladadi,Pedro3},  in which birefringence and optical rotation occur.

As well-known, the MCFJ theory is endowed with \textit{CPT} and Lorentz violation (LV), representing the \textit{CPT}-odd piece of the \textit{U}(1) gauge sector of the Standart Model Extension (SME) \cite{CFJ, Colladay, Colladay2}. The SME has been widely examined in many scenarios, such as in radiative evaluations \cite{CFJ4, CFJ4C, CFJ4D} and topological defects solutions \cite{CFJ5}.

The MCFJ electrodynamics also has a relevant connection with the axion Lagrangian \cite{Sekine, Tobar},
\begin{align}
	\mathrm{{\mathcal{L}}}=-\frac{1}{4}F^{\mu\nu}F_{\mu\nu}+\theta (\mathbf{E}\cdot \mathbf{B)},\label{Laxion}
\end{align}
where $F_{\mu\nu}=\partial_{\mu}A_{\nu}-\partial_{\nu}A_{\mu}$ is the field strength tensor and  $\theta$ is the scalar axion field. The axion is a light pseudoscalar particle that emerges in the solution of the strong CP problem in Quantum Chromodynamics \cite{Peccei}, being one of the leading dark matter candidates \cite{Preskill, Feng}. In the case the axion derivative is a constant vector, $\partial_{\mu}\theta=(k_{AF})_{\mu}$, Lagrangian (\ref{Laxion}) yields the MCFJ term,
\begin{align}
	\mathrm{{\mathcal{L}}}=-\frac{1}{4}G^{\mu\nu}F_{\mu\nu}    + \frac{1}{4}%
	\epsilon^{\mu\nu\alpha\beta}\left( k_{AF}\right)_{\mu}A_{\nu}F_{\alpha
		\beta}   ,\label{MCFJMATTER}
\end{align}
where $\left(k_{AF}\right)_{\mu}$ is the 4-vector CFJ background and $G^{\mu\nu}=\frac{1}{2}\chi^{\mu\nu\alpha\beta}F_{\alpha\beta}$ is the constitutive field strength tensor \cite{refPOST} that defines the continuous substrate where this theory is set\footnote{The 4-rank tensor, $\chi^{\mu\nu\alpha\beta}$ is the medium constitutive tensor \cite{refPOST}, which provides the electric and magnetic responses of the medium. The electric permittivity and magnetic permeability tensors are $\epsilon_{ij}\equiv \chi^{0ij0}$ and $\mu^{-1}_{lk}\equiv \frac{1}{4} \epsilon_{ijl}\chi^{ijmn}\epsilon_{mnk}$, respectively.  The usual isotropic constitutive relations are provided from isotropic polarization and magnetization. Then, it holds $\epsilon_{ij}=\epsilon  \delta_{ij}$ and $\mu^{-1}_{ij}=\mu^{-1} \delta_{ij}$, which yield the relations  $\mathbf{D}=\epsilon\mathbf{E}$, $\mathbf{H}= \mu^{-1}\mathbf{B}$.}. Lagrangian (\ref{MCFJMATTER}) describes the MFCJ electrodynamics in matter and yields the modified inhomogeneous Maxwell equations, 
\begin{align}
	\nabla\cdot\mathbf{D}  &=-\mathbf{k}_{AF}%
	\cdot\mathbf{B}  ,\label{Coulomb1}\\
	\nabla\times\mathbf{H} -\frac{\partial\mathbf{D}
	}{\partial t} &=  -  k_{AF}^{0}    \mathbf{B}+\mathbf{k}_{AF}\times
	\mathbf{E} ,\label{Amp1} 
\end{align}
where $G^{i0}=D^{i}$,  $G^{ij}=-\epsilon_{ijk}H^{k}$,  and $\quad (k_{AF})^{\mu}=(k_{AF}^{0}, \mathbf{k}_{AF})$. The homogeneous Maxwell equations are obtained from the Bianchi identity $\partial_{\mu} \tilde{F}^{\mu\nu}=0$. 

The modified Ampère's law (\ref{Amp1}) contains a chiral  magnetic current density
\begin{equation} 
	\mathbf{J}_{B}=k_{AF}^{0}\mathbf{B}, 
\end{equation}
where $k_{AF}^{0}$ plays the role of magnetic conductivity. The other additional term that appears in Ampère's law is the anomalous Hall current density, $	\mathbf{J}_{AH}=\mathbf{k}_{AF} \times\mathbf{E}$,
where the chiral vector $\mathbf{k}_{AF}$ represents the anomalous Hall conductivity. These currents have been used to investigate electromagnetic properties of matter endowed with the CME and AHE \cite{Qiu}, in connection with the description of Weyl semimetals.

The chiral magnetic effect (CME) \cite{Kharzeev1A, Fukushima, Kharzeev1C, LiKharzeev} is a macroscopic quantum effect and consists of a linear magnetic current law, ${\bf J}={\sigma}_{B}{\bf B}$, arising from an asymmetry (imbalance)
	between the number density of left- and right-handed chiral fermions, being largely addressed in many distinct contexts \cite{Schober, Vilenkin, Akamatsu, Boyarsky, Leite,Dvornikov}. Non-linear CME has also been considered in Weyl semimetals (WSMs) \cite{Burkov}, where electric and magnetic fields applied yield a current effectively proportional to $B^{2}$, that is,  ${\bf J}={\sigma}({\bf E} \cdot {\bf B}){\bf B}$ \cite{ Barnes}. 

Cold magnetized plasma has been examined in the context of chiral bi-isotropic extended constitutive relations, with the attainment of the negative refraction associated with the chirality	parameter \cite{Guo, Gao}. Chiral cold plasmas were also investigated in the scenario of the axion-like Maxwell-Carroll-Field-Jackiw (MCFJ) electrodynamics \cite{Filipe1, Filipe2}, considering the influence of the chiral magnetic current and anomalous Hall current. The properties of the propagating modes have been discussed, such as birefringence, absorption, and optical rotation, highlighting the role played by the chiral scalar factor, $k_{AF}^{0}$, and the chiral vector $\mathbf{k}_{AF}$, both engendering RP reversion.  

Chiral plasmas effects in astrophysics have also been investigated in pulsars and black holes \cite{Gorbar2}, objects surrounded by magnetospheres made of plasma. In such regions, chiral plasmas are shown to be relevant in connection with chiral anomalous processes, where the CME current  
	\begin{equation} 
		\mathbf{J}_{B}=\mu_{5}\mathbf{B}, 
		\label{MagC1}
	\end{equation}
is supposed to exist, with repercussions on the propagation of helical modes \cite{Gorbar2}. The role of the CME in the generation of strong magnetic fields in neutron stars has also been examined \cite{Maxim3}, and its relevance in astrophysical systems has been a topical issue, see Ref.~\cite{Kamala}. The chiral current (\ref{MagC1}) could also play an interesting role in the examination of a symmetrical cold electron-positron plasma, where the electron density is equal to the positron density, $n_e = n_p$, with  $m_e =m_p$. This symmetry condition makes null the nondiagonal terms of the susceptibility tensor of the magnetized cold plasma, eliminating optical activity effects \cite{Gueroult2}, opening the possibility to the current (\ref{MagC1}) playing key roles not yet examined. 

Recently, plasmas systems have been considered in experimental proposals for dark matter detection in plasmon-axion conversion \cite{Millar1}, tunable cryogenic plasmas \cite{Millar2} and also in experimental surveys for axions and dark photons in the Axion Longitudinal Plasma Haloscope (ALPHA) Consortium \cite{Millar3}. In the context of axion electrodynamics, optical properties were discussed for an axion-plasma background \cite{JMcDonald}. Coupling with the Langmuir waves was addressed in the dynamical axion scenario \cite{Tercas2}, where a new quasiparticle excitation, called axion-plasmon polariton, is identified from the modification of the Langmuir waves dispersion relation. The axion-plasmon coupling has also been considered in the shadows of rotating black holes \cite{Khodadi}.

Although conventional cold plasmas do not exhibit optical activity in the absence of an external magnetic field under simplest conditions, unusual chiral propagating modes for ionization waves (chiral streamers) in a non-magnetized plasma at sub-atmospheric-pressure condition were reported in 2015 \cite{Zou1}. Such chiral streamers and their different orientations, right- and left-handed, were recently explained through the surface electromagnetic standing wave theory \cite{Zou2}. Chiral plasma plumes have also been studied in the absence of an external magnetic field \cite{Zou3, Zou4}.

 In this work, we study the propagation of electromagnetic waves in an unmagnetized chiral (parity-odd) plasma ruled by the MCFJ electrodynamics endowed with the purely timelike chiral factor, $\left(  k_{AF}\right)^{\mu}=\left(k_{AF}^{0}, \textbf{0}\right)$, which also plays the role of chiral magnetic conductivity. This restriction is also very usual in cold dark matter research, where the space variation of the axion field can be neglected, with $\mathbf{\nabla }\theta=\mathbf{k}_{AF}=0$. The main purpose of this work is to investigate propagation and absorption zones, as well as the conditions of such behaviors. Thus, the dispersion relations and refractive indices are determined. Optical effects, such as birefringence and dichroism, are discussed and used to compare different cold chiral plasma scenarios with the magnetized conventional behavior. The chiral factor appears bringing about new properties to the unmagnetized panorama, as optical activity, negative refraction, and RP sign reversion. This paper is outlined as follows. In Sec.~\ref{Propag.TL}, we obtain the dispersion relation, refractive indices, and propagating modes for the chiral cold unmagnetized plasma under consideration. The profile of the refractive indices is examined for two possible configurations. The dispersion relations, $\omega \times k$, are also displayed and discussed. In Sec.~\ref{birefringence}, optical effects (birefringence and dichroism) are addressed in the cases examined. In Sec.~\ref{conclusion}, we summarize the results and perspectives.

{\section{Wave propagation in chiral plasma}
\label{Propag.TL} }

In this section, we derive the collective electromagnetic modes for a cold unmagnetized chiral plasma, whose modified Maxwell equations yield
	\begin{equation}
		\left[n^{2}\delta_{ij}-n^{i}n^{j}-\frac{{\varepsilon}_{ij}%
		}{\varepsilon_{0}}-i \frac{V_{0}}{\omega} \epsilon_{ikj}n^{k} \right] {E}^{j}=0,
		\label{EWE2}
	\end{equation}
with the eletrical permittivity $\varepsilon_{ij}$ given in (\ref{Epip1}) and $V_{0}=k_{AF}^{0}/\varepsilon_{0}$, the redefined timelike component of the chiral background, while $\mathbf{n}=c\mathbf{k}/\omega$ is the refractive index. In matrix form, Eq. (\ref{EWE2}) reads

\begin{widetext}
\begin{equation}
\begin{bmatrix}
n^{2}-n_{x}^{2}-P & -n_{x}n_{y}+i\left(V_{0}/\omega\right)n_{z} &
-n_{x}n_{z}-i\left(V_{0}/\omega\right)n_{y}\\
-n_{x}n_{y}-i\left(V_{0}/\omega\right)n_{z} & n^{2}-n_{y}^{2}-P & -n_{y}n_{z}+i\left(V_{0}/\omega\right)n_{x}\\
-n_{x}n_{z}+i\left(V_{0}/\omega\right)n_{y} & -n_{y}n_{z}-i\left(V_{0}/\omega\right)n_{x} & n^{2}-n_{z}^{2} -P
\end{bmatrix}
\begin{bmatrix}
\delta E_{x}\\
\delta E_{y}\\
\delta E_{z}%
\end{bmatrix}
=0,\label{timelike2}
\end{equation}
\end{widetext}
where 
\begin{equation}
P=1- \frac{\omega_{p}^{2}} {\omega^2}. \label{Pusual}
\end{equation}
For simplicity, let us consider the refractive index parallel to the magnetic field, $\mathbf{n}=n\hat{z}$, such that the latter system becomes
\begin{equation}
\begin{bmatrix}
n^{2}-P & i\left(V_{0}/\omega\right)n &
0\\
-i\left(V_{0}/\omega\right)n & n^{2}-P & 0\\
0 & 0 & -P
\end{bmatrix}
\begin{bmatrix}
\delta E_{x}\\
\delta E_{y}\\
\delta E_{z}%
\end{bmatrix}
=0, \label{timelike}
\end{equation}
whose solution stems from the null determinant condition, $\mathrm{det}[M_{ij}] =0$,
	\begin{equation}
	P \left[  \omega^{2}
	\left(  n^{2}-P\right)  ^{2}-n^2 V_{0}^{2}\right]=0,
	\label{DR1A}
	\end{equation}
the dispersion equation of the system. For longitudinal wave configuration,  $\delta\mathbf{E}=(0,0,\delta E_{z})$, a non propagating mode appears at the plasma frequency, $\omega=\omega_{p}$, associated with $P=0$. Such a longitudinal oscillation is the same one existing for a usual unmagnetized plasma, being called plasmon. Plasmon is also known in solid-state context, representing the collective mode of electrons vibrating (longitudinally) under the action of the electromagnetic field.
 
For transverse waves,   $\mathbf{n}\perp\delta\mathbf{E}$ or $\delta\mathbf{E}=(\delta E_{x},\delta E_{y},0)$, the dispersion relation (\ref{DR1A}) yields
\begin{equation}
\left(n^2-P\right)^2-n^2 \left(V_{0}/\omega\right)^2=0,
\end{equation}
also written as,
\begin{equation}
n^{2}\pm\frac{V_{0}}{\omega}n-P   =0,
\end{equation}
which provides, using the relation (\ref{Pusual}), the following refractive indices:
\begin{align}
	n_{R,M} &= - \frac{V_{0}}{2\omega} \pm \sqrt{ 1 + \left( \frac{V_{0}}{2\omega}\right)^{2} - \frac{\omega_{p}^{2}}{\omega^2}} , \label{n-R-M-indices-1} \\
	n_{L,E} &= \frac{ V_{0}}{2\omega} \pm \sqrt{ 1 + \left( \frac{V_{0}}{2\omega}\right)^{2} - \frac{\omega_{p}^{2}}{\omega^2}}. \label{n-L-E-indices-1}
	\end{align}
The refractive indices above may be real, imaginary, or complex (presenting both pieces) at some frequency ranges, where real and imaginary parts are associated with the propagation and absorption of the electromagnetic modes, respectively. 

The propagating modes associated with the refractive indices in (\ref{n-R-M-indices-1}) and (\ref{n-L-E-indices-1}) are the eigenvectors (with a null eigenvalue) obtained from the system (\ref{timelike}) and written in terms of circularly polarized waves. One obtains an LCP mode, associated with the indices $n_L$ and $n_{E}$,
	\begin{equation}
	n_{L}, n_{E}  \   \mapsto  \ \mathbf{{E}}_{LCP}=\frac{i}{\sqrt{2}}%
	\begin{bmatrix}
	1 \\ 
	i\\
	0
	\end{bmatrix}, \label{ELCP2}
	\end{equation}
and an RCP mode connected with the indices $n_R$ and $n_M$,
	\begin{equation}
	n_{R}, n_{M} \   \mapsto  \ \mathbf{{E}}_{RCP}=\frac{i}{\sqrt{2}}%
	\begin{bmatrix}
	1 \\ 
	-i\\
	0
	\end{bmatrix}.\label{ERCP2}
	\end{equation}

It is easy to notice that the indices $n_R$, $n_{E}$, given by Eqs. (\ref{n-R-M-indices-1}) and (\ref{n-L-E-indices-1}), share the same standard cutoff frequency, $\omega_{p}$, while the refractive indices $n_{L}$ and  $n_{M}$ have no real root. It is important to notice that the radicand,
	\begin{equation}
	R\left(\omega\right)= 1 + \left( \frac{V_{0}}{2\omega}\right)^{2} - \frac{\omega_{p}^{2}}{\omega^2}, \label{radicando}
	\end{equation}
in Eqs. (\ref{n-R-M-indices-1}) and (\ref{n-L-E-indices-1}), has a single positive root, given by
	\begin{equation}
	\omega_{\mathit{rad}}=\sqrt{\omega_{p}^{2}-V_{0}^{2}/4},
	\end{equation}
which, when real or imaginary, defines two main scenarios, introduced below.
\begin{itemize}
	\item Scenario 1: for	
	\begin{equation}
		V_{0}^{2}/4<\omega_{p}^{2},\label{condition1}
	\end{equation} 
the root $\omega_{\mathit{rad}}$ is real, and the factor $R\left(\omega\right)$ can be negative for $\omega<\omega_{\mathit{rad}}$ or positive for $\omega>\omega_{\mathit{rad}}$. In this case, the indices (\ref{n-R-M-indices-1}) and (\ref{n-L-E-indices-1}) are complex for $\omega<\omega_{\mathit{rad}}$ and real for $\omega>\omega_{\mathit{rad}}$.
	\item Scenario 2: for		
	\begin{equation}
		V_{0}^{2}/4>\omega_{p}^{2},\label{condition2}
	\end{equation}
the root $\omega_{\mathit{rad}}$ is imaginary, and the factor $R\left(\omega\right)$ is always positive,  implying real indices for any frequency.
\end{itemize}

In the following, the behavior of the refractive indices (\ref{n-R-M-indices-1}) and (\ref{n-L-E-indices-1}) is examined and illustrated, considering the conditions (\ref{condition1}) and (\ref{condition2}). In the absence of the chiral factor, $V_{0}=0$, the indices (\ref{n-R-M-indices-1}) and (\ref{n-L-E-indices-1}) reduce to the unmagnetized conventional expressions,
\begin{equation}
	\tilde{n}_{\pm} =  \pm \sqrt{ 1 - \frac{\omega_{p}^{2}}{\omega^2}}, \label{n_usual-indices} \\
\end{equation}
which will be used as the standard comparison for indicating the chiral factor effects.

\subsection{About the index $n_{R}$ \label{secNR}} 

\begin{figure}[h]
	\centering
	\includegraphics[scale=0.6]{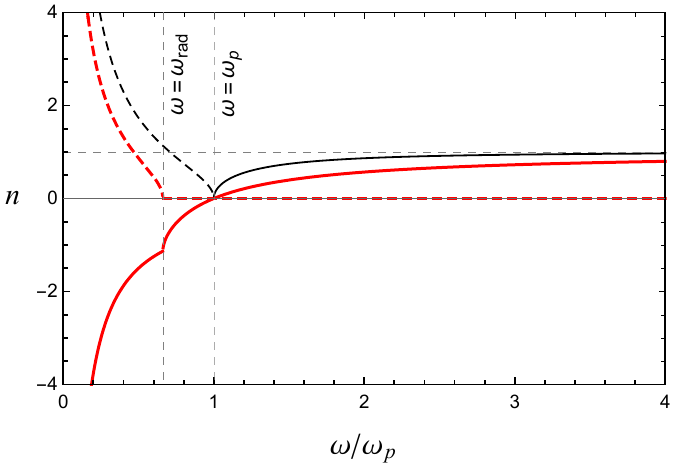}  \caption{Index of refraction 
			$n_{R}$ under the condition (\ref{condition1}). The dashed red (black) line corresponds to the imaginary piece of $n_{R}$ ($\tilde{n}_+$), while the solid red (black) line represents the real piece of $n_{R}$ ($\tilde{n}_+$), where the index $\tilde{n}_+$ is given in \eqref{n_usual-indices}. The scalar chiral parameter induces a new propagation window of negative refraction near the plasma frequency. Here, $V_{0}=(3/2)\omega_{p}$, and $\omega_{p}=1$~$\mathrm{rad}$~$s^{-1}$.}
	\label{nRfig}%
\end{figure}

\begin{figure}[h]
	\centering
	\includegraphics[scale=0.6]{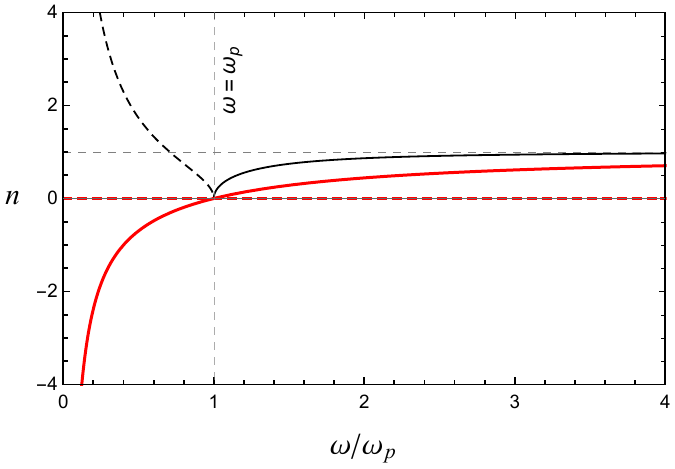}  \caption{Index of refraction 
			$n_{R}$ under the condition (\ref{condition2}). The dashed red (black) line corresponds to the imaginary piece of $n_{R}$ ($\tilde{n}_+$), while the solid red (black) line represents the real piece of $n_{R}$ ($\tilde{n}_+$).  The scalar chiral parameter induces a propagation window of negative refraction for $0<\omega<\omega _{p}$. Here, $V_{0}=(5/2)\omega_{p}$, and $\omega_{p}=1$~$\mathrm{rad}$~$s^{-1}$.}
	\label{nRfig2}%
\end{figure}

We now discuss some aspects of the index $n_{R}$, which has a root at $\omega=\omega _{p}$. The behavior of $n_{R}$ in terms of the dimensionless parameter $\omega/\omega_{c}$ is illustrated in Figs.~\ref{nRfig} and \ref{nRfig2}, which show the real and imaginary pieces of the refractive index $n_{R}$ under the conditions (\ref{condition1}) and (\ref{condition2}), respectively.


For the condition (\ref{condition1}), we point out:
\begin{enumerate}
	[label=(\roman*)]
	
	\item  For $\omega\rightarrow 0$, the real piece of $n_{R}$ tends to infinity (negatively), $\mathrm{Re}[n_{R}] \rightarrow -\infty$, while its imaginary part tends to infinity (positively),  $\mathrm{Im}[n_{R}]\rightarrow \infty$. This behavior differs from the usual case, where $\mathrm{Re}[\tilde{n}_+] =0$ and $\mathrm{Im}[\tilde{n}_+] \rightarrow \infty$ as $\omega\rightarrow 0$. See black lines in Fig. \ref{nRfig}.
	
	\item For $0<\omega<\omega _{\mathit{rad}}$, it occurs a negative refractive index zone with absorption, where $\mathrm{Re}[n_{R}]<0$ and $\mathrm{Im}[n_{R}]\ne0$, as shown in Fig.~\ref{nRfig}. In the usual case, this negative refraction region does not exist.
	
	\item For $\omega _{\mathit{rad}}<\omega<\omega_{p}$, an attenuation-free propagation zone with negative refraction is opened, that is, $\mathrm{Re}[n_{R}]<0$ and $\mathrm{Im}[n_{R}]=0$.
	
	\item For $\omega>\omega_{p}$, $n_{R}$ is always real and positive, $\mathrm{Re}[n_{R}]>0$ and $\mathrm{Im}[n_{R}]=0$, corresponding to a propagation zone, with $n_{R} \rightarrow1$ in the high-frequency limit.  
	
\end{enumerate}

On the other hand, under the condition (\ref{condition2}), only propagation zones are reported. The first one occurs for the range $0<\omega<\omega _{p}$ and is marked with negative refraction ($\mathrm{Re}[n_{R}]<0$ and $\mathrm{Im}[n_{R}]=0$). The second one is defined for $\omega>\omega_{p}$, where $\mathrm{Re}[n_{R}]$ increases monotonically in the high-frequency limit, as illustrated in Fig.~\ref{nRfig2}.

\subsection{About the index $n_{M}$ \label{secNM}}

\begin{figure}[h]
	\centering
	\includegraphics[scale=0.60]{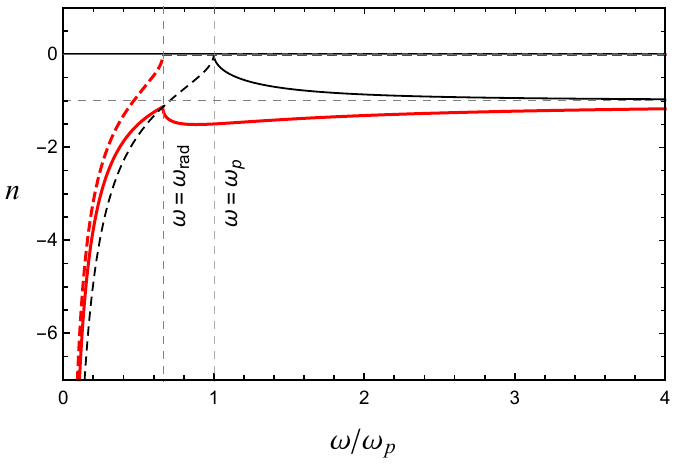}  \caption{Index of refraction $n_{M}$ under the condition (\ref{condition1}). The dashed red (black) line corresponds to the imaginary piece of $n_{M}$ ($\tilde{n}_-$), while the solid red (black) line represents the real piece of $n_{M}$ ($\tilde{n}_-$).  Here, $V_{0}=(3/2)\omega_{p}$, and $\omega_{p}=1$~$\mathrm{rad}$~$s^{-1}$.}
	\label{nMfig}%
\end{figure}

The index $n_{M}$, given in Eq. (\ref{n-R-M-indices-1}) and plotted in Fig.~\ref{nMfig}, is always negative, having no real root, as it happens in the magnetized chiral plasma as well (see Ref.~\cite{Filipe1}). Under the condition (\ref{condition1}), it holds

\begin{figure}[h]
	\centering
	\includegraphics[scale=0.6]{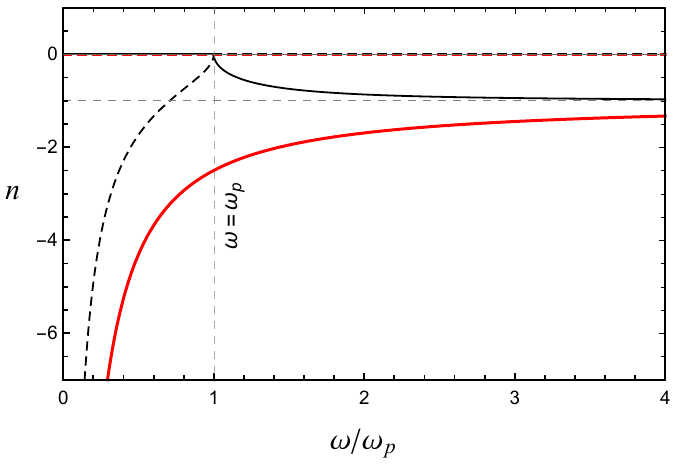}  \caption{Index of refraction $n_{M}$ under the condition (\ref{condition2}). The dashed red (black) line corresponds to the imaginary piece of $n_{M}$ ($\tilde{n}_-$), while the solid red (black) line represents the real piece of $n_{M}$ ($\tilde{n}_-$).  Here, $V_{0}=(5/2)\omega_{p}$, and $\omega_{p}=1$~$\mathrm{rad}$~$s^{-1}$.}
	\label{nMfig2}%
\end{figure}

\begin{enumerate}
	[label=(\roman*)]
	
	\item  For $\omega\rightarrow 0$, the real and imaginary pieces of $n_{M}$ diverges (negatively), $\mathrm{Re}[n_{M}], \mathrm{Im}[n_{M}] \rightarrow -\infty$. For $0<\omega<\omega _{\mathit{rad}}$, the index  $n_{M}$ presents a negative refraction zone with absorption, that is, $\mathrm{Re}[n_{M}]<0$ and $\mathrm{Im}[n_{M}]\ne0$, as shown in Fig.~\ref{nMfig}. This behavior contrasts with the one of the unmagnetized conventional index, $\tilde{n}_-$,  whose real piece is null (it does not propagate). See the dashed black line in Fig.~\ref{nMfig}.

	\item For $\omega>\omega _{\mathit{rad}}$, it occurs a propagation zone with negative refraction, $\mathrm{Re}[n_{M}]<0$ and $\mathrm{Im}[n_{M}]=0$, expanding the full propagation zone that, in the usual case, is defined for $\omega>\omega _{p}$. 
	
	\item For $\omega\rightarrow \infty$, $n_{M} \rightarrow -1$, the same behavior of the usual case in the high-frequency limit.
	 
\end{enumerate}

Considering the condition (\ref{condition2}), the index $n_{M}$ is always real and negative for $\omega>0$, with a monotonically decreasing magnitude, as shown in Fig.~\ref{nMfig2}, which represents a propagation zone with negative refraction for all frequencies. 

\subsection{About the index $n_{L}$ \label{secNL}}
\begin{figure}[h]
	\centering
	\includegraphics[scale=0.55]{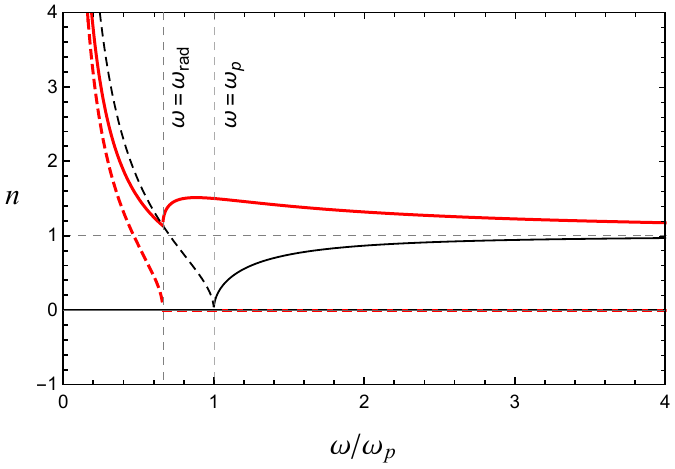}  \caption{Index of refraction $n_{L}$ under the condition (\ref{condition1}). The dashed red (black) line corresponds to the imaginary piece of $n_{L}$ ($\tilde{n}_+$), while the solid red (black) line represents the real piece of $n_{L}$ ($\tilde{n}_+$).  Here, $V_{0}=(3/2)\omega_{p}$, and $\omega_{p}=1$~$\mathrm{rad}$~$s^{-1}$.}
	\label{nLfig}%
\end{figure}

The index $n_{L}$ is always positive and has no real root. It is the mirror image of the index $n_{M}$ in relation to the abscissas axis. Under the condition (\ref{condition1}), its general behavior is represented in Fig.~\ref{nLfig} and detailed below.
\begin{enumerate}
	[label=(\roman*)]
	
	\item  For $\omega\rightarrow 0$,  $\mathrm{Re}[n_{L}],\mathrm{Im}[n_{L}] \rightarrow +\infty$. For $0<\omega<\omega _{\mathit{rad}}$, the index is complex,  $\mathrm{Re}[n_{L}]>0$ and $\mathrm{Im}[n_{L}]\ne0$, as highlighted in Fig.~\ref{nLfig}. This partial propagating behavior differs from the standard unmagnetized case, where there is only absorption in this range.
	
	\item For $\omega>\omega _{\mathit{rad}}$, there appears a full propagation zone, with $\mathrm{Re}[n_{L}]>0$ and $\mathrm{Im}[n_{L}]=0$, enlarged in comparison with the usual case, which begins for $\omega>\omega _{p}$.
	
	\item For $\omega\rightarrow \infty$, $n_{L} \rightarrow 1$, a similar behavior to the usual case in this limit.
\end{enumerate}

\begin{figure}[h]
	\centering
	\includegraphics[scale=0.55]{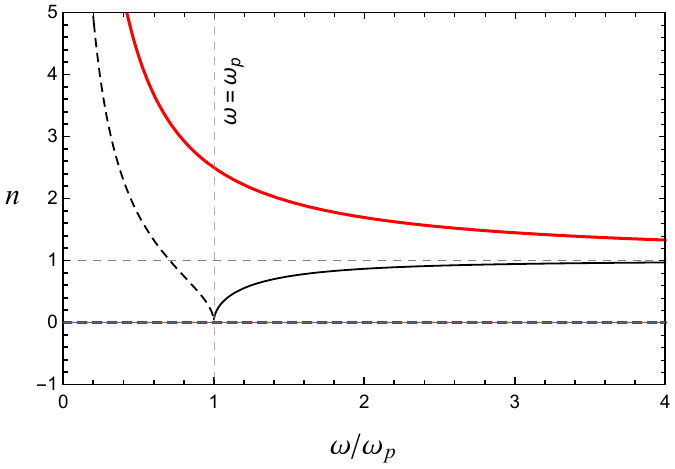}  \caption{Index of refraction $n_{L}$ under the condition (\ref{condition2}). The dashed red (black) line corresponds to the imaginary piece of $n_{L}$ ($\tilde{n}_+$), while the solid red (black) line represents the real piece of $n_{L}$ ($\tilde{n}_+$).  Here, $V_{0}=(5/2)\omega_{p}$, and $\omega_{p}=1$~$\mathrm{rad}$~$s^{-1}$.}
	\label{nLfig2}%
\end{figure}

Under the condition (\ref{condition2}) the refractive index $n_{L}$ is depicted in Fig.~\ref{nLfig2}, being always positive for $\omega>0$, with $\mathrm{Re}[n_{L}]>0$  and $\mathrm{Im}[n_{L}]=0$. It is the mirror image of the index $n_{M}$ in relation to the frequency axis, see Fig. \ref{nMfig2}, corresponding to a propagating mode for all frequencies.

\subsection{About the index $n_{E}$ \label{secNE}}

\begin{figure}[h]
	\centering
	\includegraphics[scale=0.55]{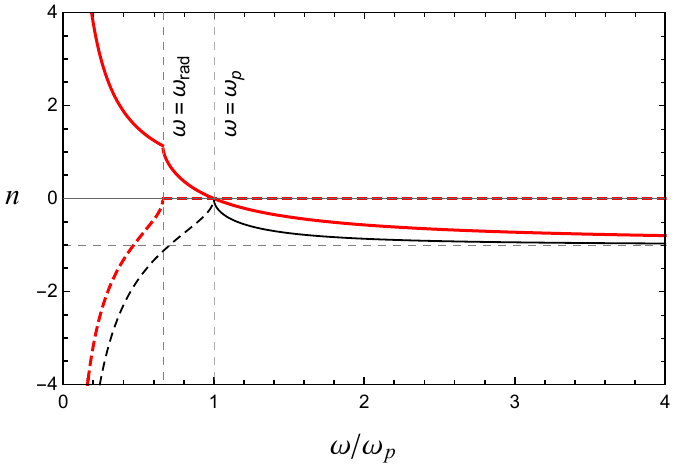}  \caption{Index of refraction $n_{E}$ under the condition (\ref{condition1}). The dashed red (black) line corresponds to the imaginary piece of $n_{E}$ ($\tilde{n}_-$), while the solid red (black) line represents the real piece of $n_{E}$ ($\tilde{n}_-$).  Here, $V_{0}=(3/2)\omega_{p}$, and $\omega_{p}=1$~$\mathrm{rad}$~$s^{-1}$.}
	\label{nEfig1}%
\end{figure}

The index $n_{E}$, given in Eq. (\ref{n-L-E-indices-1}), has root at $\omega=\omega_{p}$.  Under the condition (\ref{condition2}), it presents the following features:

\begin{enumerate}
	[label=(\roman*)]
	\item  For $\omega\rightarrow 0$, one has $\mathrm{Re}[n_{E}] \rightarrow \infty$ and $\mathrm{Im}[n_{E}]\rightarrow -\infty$, which differs from the  usual case, $\mathrm{Re}[\tilde{n}_-]=0$ as $\omega\rightarrow 0$. For $0<\omega<\omega _{\mathit{rad}}$, the index $n_{E}$ is complex, as highlighted in Fig.~\ref{nEfig1}.
	
	\item For $\omega _{\mathit{rad}}<\omega<\omega_{p}$, the index is real, positive, and decreasing with $\omega$, representing a propagation zone that does not exist in standard case. See Fig.~\ref{nEfig1}
	
	\item For $\omega> \omega_{p}$, the index $n_{E}$ becomes negative, defining a propagation zone with negative refraction, where its magnitude progressively increases to the asymptotic value, $n_{E} \rightarrow -1$, in a very similar way of the usual case.

\end{enumerate}

\begin{figure}[h]
	\centering
	\includegraphics[scale=0.55]{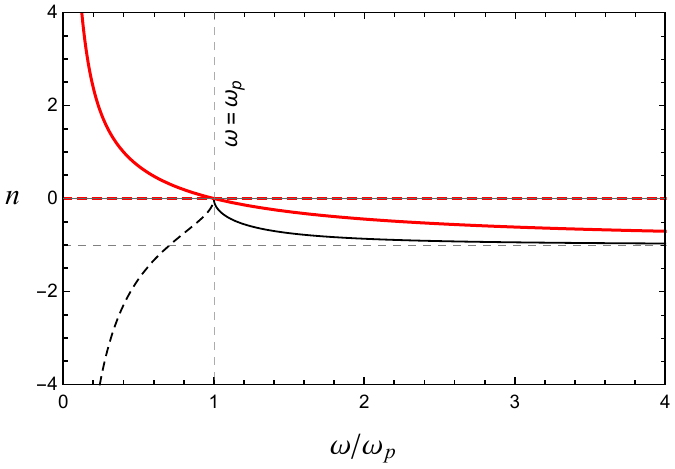}  \caption{Index of refraction $n_{E}$ under the condition (\ref{condition2}). The dashed red (black) line corresponds to the imaginary piece of $n_{E}$ ($\tilde{n}_-$), while the solid red (black) line represents the real piece of $n_{E}$ ($\tilde{n}_-$).  Here, $V_{0}=(5/2)\omega_{p}$, and $\omega_{p}=1$~$\mathrm{rad}$~$s^{-1}$.}
	\label{nEfig2}%
\end{figure}

On the other hand, for the condition (\ref{condition2}), there is no absorption, since $n_{E}$ is real for any frequency, see Fig.~\ref{nEfig2}. In this case,   $\mathrm{Re}[n_{E}]>0$ for $\omega<\omega_{p}$, corresponding to a propagation zone that appears in the place of the full usual absorbing zone. Furthermore, for $\omega>\omega_{p}$, $\mathrm{Re}[n_{E}]<0$, and there is a propagation zone with negative refraction, presenting a similar behavior to the usual case.

\subsection{\label{section-helicons}Low-frequency modes}

Low-frequency modes are usually defined in magnetized plasmas, considering the regime
\begin{align}
\omega\ll \omega_{p}, \quad \omega_{c}\ll\omega_{p}, \quad  \omega\ll \omega_{c}, \label{helicon-frequency-regime}
\end{align}
which provides a well-known RCP helicon mode, described by the following refractive index:
\begin{align}
n_{-} &= \omega_{p} \sqrt{\frac{1}{\omega \omega_{c}}}, \label{helicons-12}
\end{align}
where $\omega_{c}$ is the cyclotron frequency for a charged particle in a magnetic field. Helicons are RCP modes that propagate at very low frequencies and along the magnetic field axis. See Ref. \cite{Bittencourt} {(Chapter 9)}, and Ref. \cite{chapter-8} {(Chapter 8)} for basic details. In a non-magnetized plasma, the low-frequency limit is defined by only one condition, 
\begin{align}
\omega\ll \omega_{p}, \label{helicon-frequency-regime2}
\end{align}
which implemented on the refractive indices (\ref{n-R-M-indices-1}) and (\ref{n-L-E-indices-1}), yields 
\begin{align}
	\bar{n}_{R,M}&=-\frac{ V_{0}}{2\omega} \pm \frac{1}{2\omega}\sqrt{V_{0}^{2}-4\omega_{p}^2}, \label{helicons-15} \\
	\bar{n}_{L,E} &=\frac{ V_{0}}{2\omega} \pm \frac{1}{2\omega}\sqrt{V_{0}^{2}-4\omega_{p}^2}, \label{helicons-16}
	\end{align}
where we have used the ``bar'' notation to indicate the helicons quantities.
The indices (\ref{helicons-15}) and (\ref{helicons-16}) indicate the existence of both RCP and LCP helicons, which propagate in the low-energy regime, due to the presence of the chiral timelike component, $V_{0}$.
These helicon modes are dispersive, with magnitude going with the inverse of the frequency, 
\begin{itemize}
	\item Under the condition (\ref{condition1}), $V_{0}^{2}/4<\omega_{p}^2$, the indices (\ref{helicons-15}) and (\ref{helicons-16}) become complex implying lossy behavior. 
	\item Under the condition (\ref{condition2}), $V_{0}^{2}/4>\omega_{p}^2$, the indices $\bar{n}_{R, M}$ and $\bar{n}_{L, E}$ are real, being associated with propagating dispersive helicon modes. Moreover, there occurs negative refraction, associated with $\bar{n}_{R}$, $\bar{n}_{M}$, and $\bar{n}_{E}$.
\end{itemize}

In the context of the magnetized plasmas in chiral MCFJ electrodynamics, addressed in Ref.~\cite{Filipe1}, the helicons indices $\bar{n}_{R, E}$ display a constant and nondispersive behavior, 
	differently of the indices (\ref{helicons-15}) and (\ref{helicons-16}).  Another important point is that the indices (\ref{helicons-15}) and (\ref{helicons-16}) represent low-frequency propagating modes (RCP and LCP) with different refractive indices. This scenario differs from the magnetized chiral plasmas theory, where different low-frequency modes associated with the same index $\bar{n}_{R, E}$ occur, as seen in Ref. \cite{Filipe1} (Section IV.E).

\subsection{Dispersion relations behavior}

\begin{figure}[h]
	\centering
	\includegraphics[scale=0.39]{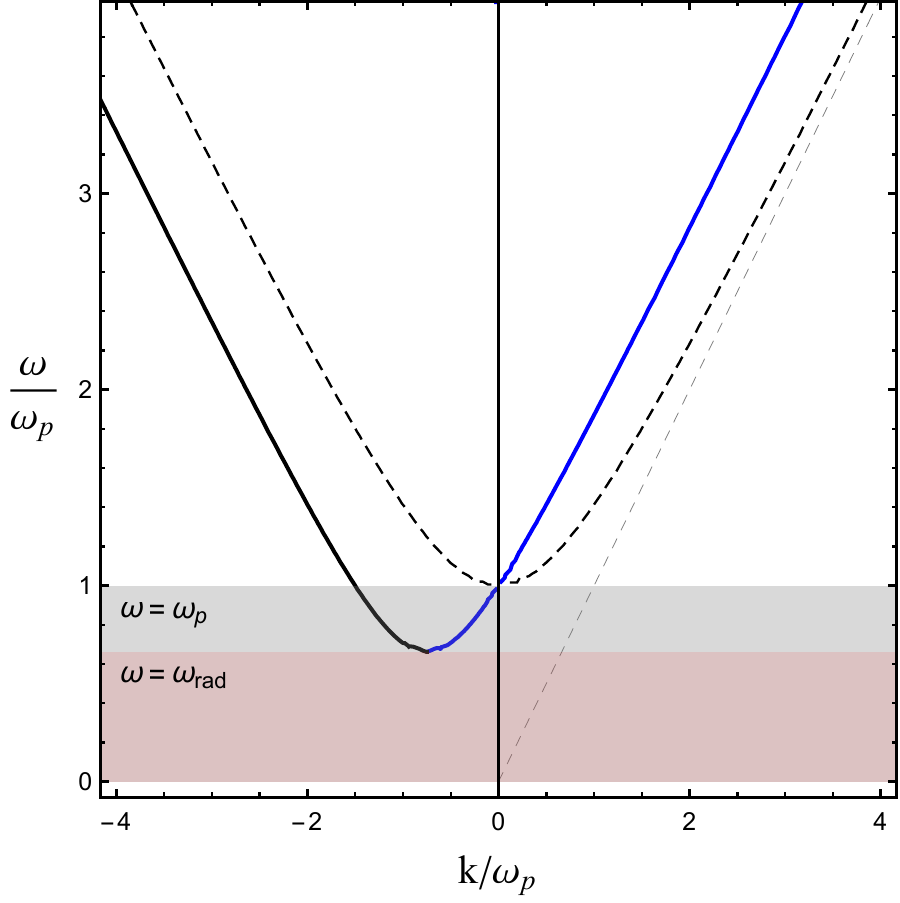}  \caption{Dispersion relations related to refractive indices $n_{R}$ (solid blue line) and $n_{M}$ (solid black line), under the  condition (\ref{condition1}). The dashed black line represents the indices of the standard case, $\tilde{n}_{\pm}$. The highlighted area in $n_{R, M}$ red (gray plus red) indicates the absorption zone for $n_{R, M}$ ($\tilde{n}_{\pm}$). Here, we have used $V_{0}=(3/2)\omega_{p}$, with $\omega_{p}=1$~$\mathrm{rad}$~$s^{-1}$.}
	\label{oxk_nr_1}%
\end{figure}

\begin{figure}[h]
	\centering
	\includegraphics[scale=0.40]{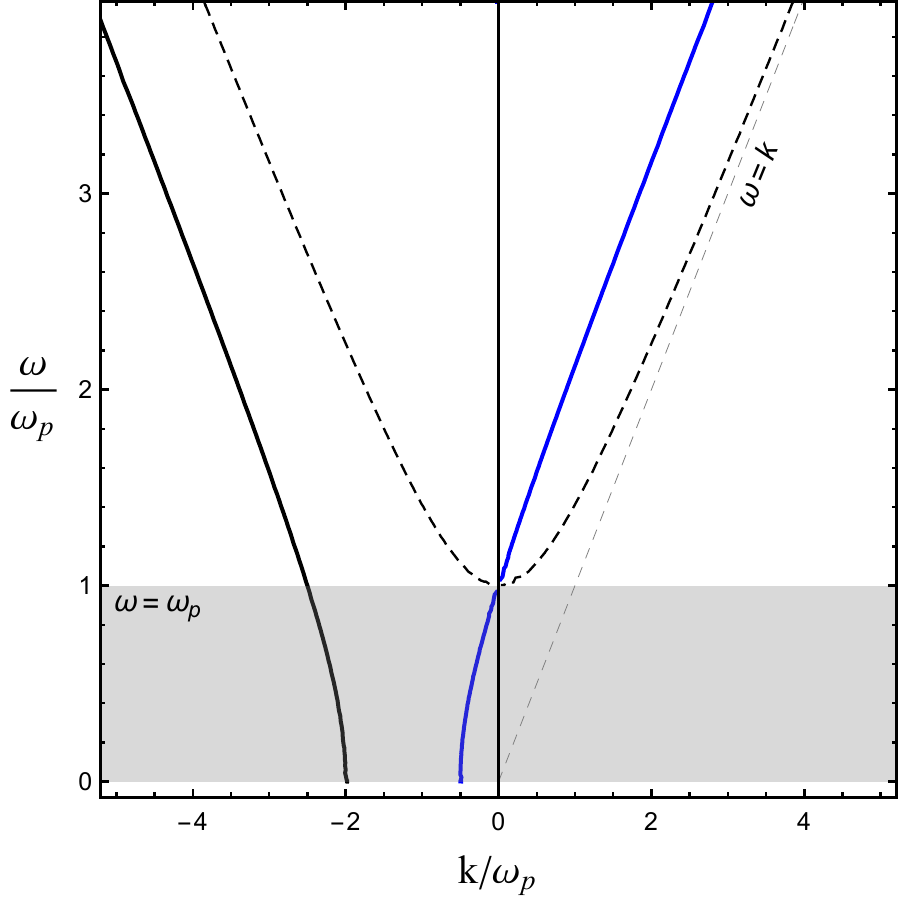}  \caption{Dispersion relations related to refractive indices $n_{R}$ (solid blue line) and $n_{M}$ (solid black line), under the  condition (\ref{condition2}). The dashed black line represents the indices of the standard case, $\tilde{n}_{\pm}$. The
			highlighted area in gray indicates the absorption zone for $\tilde{n}_{\pm}$. The plot also reveals the absence of lossy region for $n_{R, M}$. Here, we have used $V_{0}=(5/2)\omega_{p}$, with $\omega_{p}=1$~$\mathrm{rad}$~$s^{-1}$.}
	\label{oxk_nr_2}%
\end{figure}

In this section, we present a graph representation of the dispersion relations associated with the obtained circular modes in dimensionless plots of frequency $\times$ wave vector, that is, $\left(\omega/\omega_{p}\right) \times \left(k/\omega_{p}\right)$. Starting from the indices $n_{R}$ and $n_{M}$,  under the condition (\ref{condition1}), we illustrate the dispersion relations in Fig.~\ref{oxk_nr_1}. The propagation occurs for $\omega>\omega_{\mathit{rad}}$, while absorption takes place for $\omega<\omega_{\mathit{rad}}$. Since $\omega_{\mathit{rad}}<\omega_{p}$, we observe a reduced absorption zone for $n_{R}$ and $n_{M}$ (the pale pinked zone) in relation to the one of the usual indices ($\tilde{n}_{\pm}$). For $\omega_{\mathit{rad}}<\omega<\omega_{p}$, the unusual negative refraction ($k<0$) region of $n_{R}$ becomes evident. Under the condition (\ref{condition2}), the indices $n_{R}$ and $n_{M}$ are entirely real, and the associated modes propagate for any frequencies, as shown in Fig.~\ref{oxk_nr_2}. As clearly shown, the index $n_{R}$ (blue line) becomes negative for $0<\omega<\omega_{p}$.

\begin{figure}[h]
	\centering
	\includegraphics[scale=0.33]{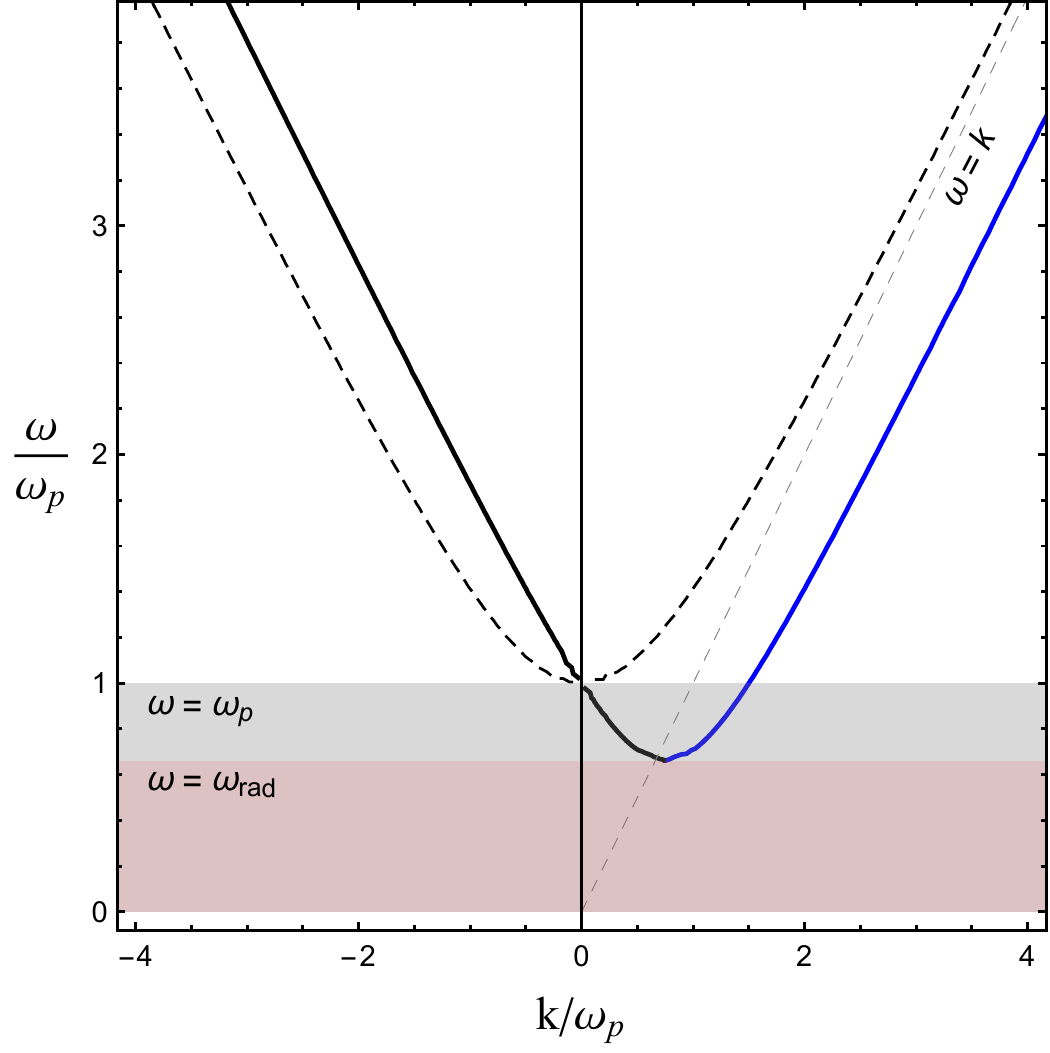}  \caption{Dispersion relations related to refractive indices $n_{L}$ (solid blue line) and $n_{E}$ (solid black line) under the  condition (\ref{condition1}). The dashed black line represents the indices of the standard case, $\tilde{n}_{\pm}$. The highlighted area in red (gray and red) indicates the absorption zone for $n_{R, M}$ ($\tilde{n}_{\pm}$). Here, we have used $V_{0}=(3/2)\omega_{p}$, with $\omega_{p}=1$~$\mathrm{rad}$~$s^{-1}$.}
	\label{oxk_nl_1}%
\end{figure}
\begin{figure}[h]
	\centering
	\includegraphics[scale=0.33]{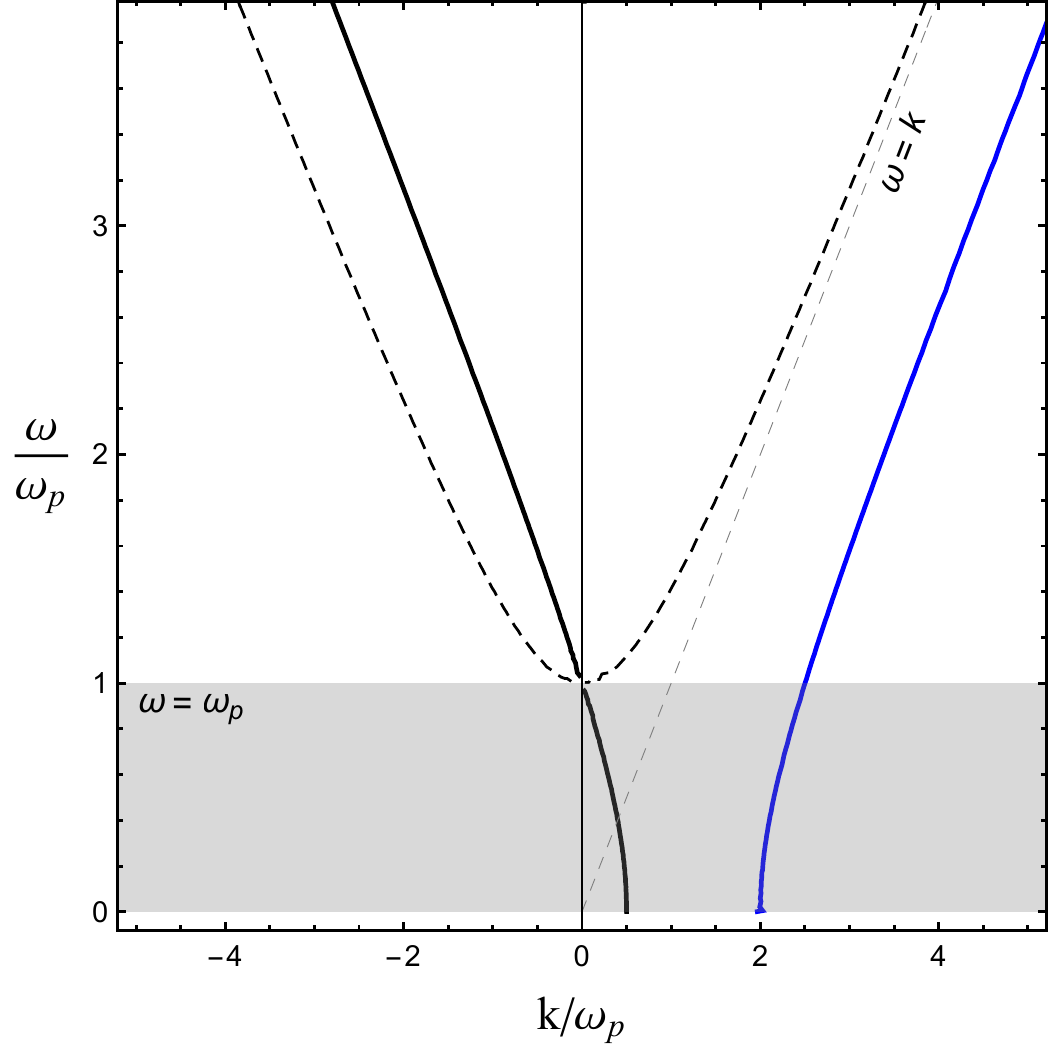}  \caption{Dispersion relations related to refractive indices $n_{L}$ (solid blue line) and $n_{E}$ (solid black line) under the  condition (\ref{condition2}). The dashed black line represents the indices of the standard case, $\tilde{n}_{\pm}$. The
			highlighted gray area represents the absorption zone for$\tilde{n}_{\pm}$. The plot also reveals the absence of a lossy region for $n_{L, E}$. Here, we have used $V_{0}=(5/2)\omega_{p}$, with $\omega_{p}=1$~$\mathrm{rad}$~$s^{-1}$.}
	\label{oxk_nl_2}%
\end{figure}

Negative refraction also arises in the chiral timelike MCFJ magnetized plasma \cite{Filipe1}, while it is not observed in the spacelike MCFJ magnetized case \cite{Filipe2}. Thus, such a negative dispersion is a feature of timelike MCFJ electrodynamics.

The dispersion relations associated with $ n_{L}$ and $n_{E}$, under the condition (\ref{condition1}), are depicted in Fig.~\ref{oxk_nl_1}. Absortion and propagation zones occurs for $\omega<\omega_{\mathit{rad}}$ and $\omega>\omega_{\mathit{rad}}$, respectively. The scenario of the shortened absorption zone is similar to the one of Fig.~\ref{oxk_nr_1}, but with the index $n_{E}$ exhibiting negative refraction in the place of  $n_{R}$. On the other hand, under the condition (\ref{condition2}), there is no absorption, and the propagation occurs for $\omega>0$, as illustrated in Fig.~\ref{oxk_nl_2}, which finds similarity with the plot of Fig.~\ref{oxk_nr_2}.

\section{Birefringence, rotatory power and dichroism \label{birefringence}}

Optical activity usually manifests as a rotation of the polarization vector of the propagating wave. Such a phenomenon, in the case of circularly polarized modes, is called circular birefringence. It occurs when the RCP and LCP modes propagate at different phase velocities. For the indices (\ref{n-R-M-indices-1}) and (\ref{n-L-E-indices-1}), the corresponding phase velocities,  $v_{R}=1/n_{R}$, $v_{L}=1/n_{L}$, $v_{E}=1/n_{E}$ and $v_{M}=1/n_{M}$, are distinct, and engender birefringence that is  measure in terms of the rotatory power, 
	\begin{equation}
		\delta=-\frac{\omega}{2}\left(\mathrm{Re}\left[n_{LCP}\right]-\mathrm{Re}\left[n_{RCP}\right]\right),
	\end{equation}
where $n_{LCP}$ and $n_{RCP}$ are the refractive indices for different circular polarizations.
The rotatory power (RP) is useful to the optical characterization of matter, including crystals \cite{Dimitriu, Birefringence1, Pajdzik2, Geday}, organic compounds \cite{Barron2, Xing-Liu}, graphene phenomena at terahertz band \cite{Poumirol}, chiral metamaterials \cite{Woo, Mun}, chiral semimetals \cite{Dey-Nandy}, and in the determination of the rotation direction of pulsars \cite{Gueroult2}. An interesting review of chiral optical phenomena is found in Ref.~\cite{Mun}, which presents a comprehensive discussion of optical features of chiral media, chiral particles, and chiral electromagnetic fields.

\subsection{Rotatory power \label{secRP}}

We write the rotatory power for the present chiral plasma considering the refractive indices $n_{L}$, $n_{E}$, associated with the LCP wave, and the indices $n_{R}$, $n_{M}$, associated to the RCP wave. Starting from the indices $n_{L}$ and $n_{R}$, the corresponding RP,
\begin{equation}
	\delta_{LR}=-\frac{\omega}{2}\left(\mathrm{Re}[n_{L}]-\mathrm{Re}[n_{R}]\right), 
\end{equation}
yields simply a constant (non-dispersive) RP, namely
	\begin{equation}
	\delta_{LR}=-\frac{\omega}{2}\mathrm{Re}\left[V_{0}/\omega\right] =-\frac{V_{0}}{2}, \label{rptl}
	\end{equation}
induced by the presence of the chiral parameter $V_{0}$.

\begin{figure}[H]
	\centering
	\includegraphics[scale=0.55]{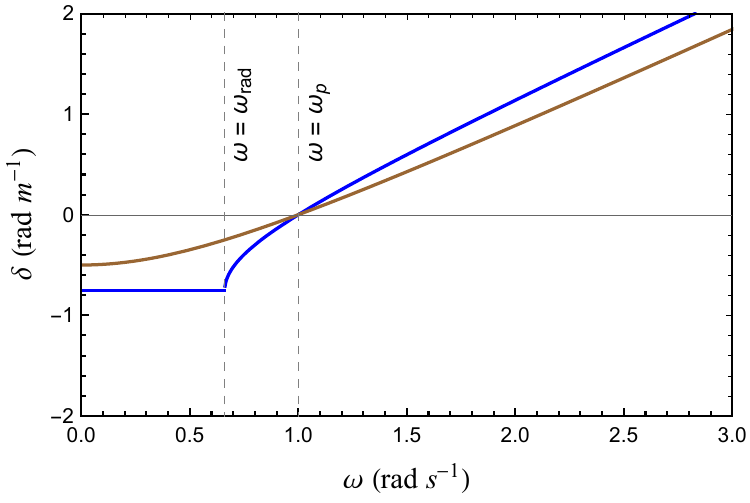}
	\caption{The rotatory power (\ref{rpeq2}) defined for the refractive indices $n_{E}$ and $n_{R}$. It is represented by the solid blue line for the condition (\ref{condition1}) and $V_{0}=(3/2)\omega_{p}$, and by the brown line for the condition (\ref{condition2}) and $V_{0}=(5/2)\omega_{p}$. Here, we have used $\omega_{p}=1$~$\mathrm{rad}$~$s^{-1}$.}
	\label{RP_nEnR}
\end{figure}

Considering now the refractive indices $n_{E}$ and $n_{R}$, the rotatory power is written as
	\begin{align}
	\delta_{ER}=-\frac{\omega}{2}\left(\mathrm{Re}[n_{E}]-\mathrm{Re}[n_{R}]\right), \\
	\delta_{ER}=-\frac{\omega}{2}\mathrm{Re}\left[V_{0}/\omega-2\sqrt{R\left(\omega\right)}\right], \label{rpeq2} 
	\end{align}	
which is depicted in Fig.~\ref{RP_nEnR}. Under the condition (\ref{condition1}), one has $R\left(\omega\right)<0$ for $\omega<\omega_{\mathit{rad}}$, so that the term $\sqrt{R\left(\omega\right)}$ is pure imaginary in this range.  The RP (\ref{rpeq2}) then coincides with the one (\ref{rptl}) for $\omega<\omega_{\mathit{rad}}$. For $\omega>\omega_{\mathit{rad}}$ the term $\sqrt{R\left(\omega\right)}$ becomes real and starts to contribute (positively) to the RP, as seen in Fig.~\ref{RP_nEnR}. It is also worth mentioning that the RP (\ref{rpeq2}) undergoes a sign reversal at the plasma frequency $\omega_{p}$. Moreover, notice that such an increasing RP is a consequence of the $n_E$ negative refraction zone.

The behavior of the RP (\ref{rpeq2}) under the condition (\ref{condition2}) is illustrated by the brown line in Fig. \ref{RP_nEnR}. In this case, one has $R\left(\omega\right)>0$ for any frequency, so the root $\sqrt{R\left(\omega\right)}$ contributes to the RP for $\omega>0$, yielding the softly increasing profile of Fig. \ref{RP_nEnR}. The sign reversal behavior at $\omega_{p}$ is not altered since it occurs for $\omega=\omega_{p}$. 

A rotatory power reversion is not usual in cold plasma theory. However, it is observed in rotating plasmas \cite{Gueroult} and in magnetized chiral cold plasmas ruled by the MCFJ electrodynamics \cite{Filipe1, Filipe2}. Such a reversion is also reported in graphene systems \cite{Poumirol} and bi-isotropic dielectrics supporting chiral magnetic current \cite{PedroPRB}.

\begin{figure}
	\centering
	\includegraphics[scale=0.55]{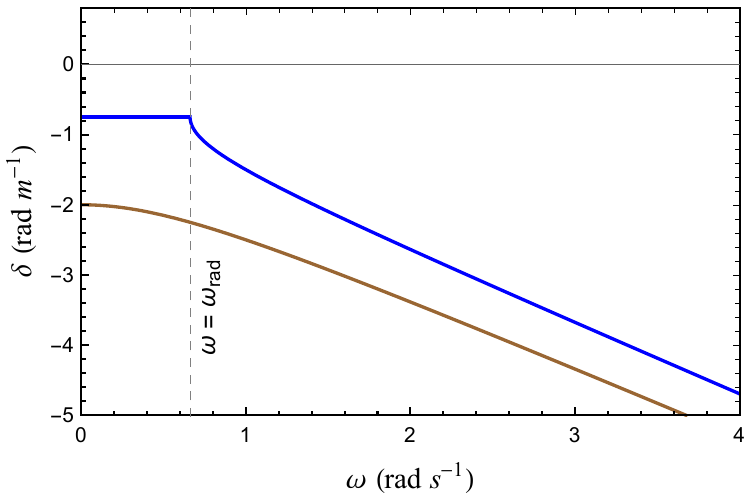}
	\caption{The rotatory power (\ref{rpeq2B}) defined for the refractive indices $n_{L}$ and $n_{M}$. It is represented by the solid blue line for the condition (\ref{condition1}) and $V_{0}=(3/2)\omega_{p}$, and by the brown line for the condition (\ref{condition2}) and $V_{0}=(5/2)\omega_{p}$. Here, we have used $\omega_{p}=1$~$\mathrm{rad}$~$s^{-1}$.}
	\label{RP_nLnM}
\end{figure}

The RP can also be evaluated for the refractive indices $n_{L}$ and $n_{M}$, yielding
\begin{equation}
	\delta_{LM}=-\frac{\omega}{2}\mathrm{Re}\left[V_{0}/\omega+2\sqrt{R\left(\omega\right)}\right].\label{rpeq2B} 
\end{equation}	
that is ploted in Fig.~\ref{RP_nLnM}, for the conditions (\ref{condition1}) and (\ref{condition2}). As noticed, the RP is always negative, with no sign reversal.

\subsection{Dichroism coefficients \label{secDC} }

The dichroism circular effect occurs when circularly polarized modes undergo absorption at different degrees in the zones where the refractive indices are complex. Taking into account the imaginary pieces of the refractive indices, the circular dichroism for LCP and RCP waves is expressed in terms of the coefficient

\begin{equation}
\delta_{d} =-\frac{\omega }{2}\left( \mathrm{Im}[n_{LCP}]-\mathrm{Im}[n_{RCP}]\right).  \label{dicro_eqtl}
\end{equation}

As already seen, under the condition (\ref{condition1}) the refractive indices $n_{L}$ and $n_{R}$ are complex for $\omega<\omega_{\mathit{rad}}$ and there occurs absorption. The dichroism coefficient, however,
	\begin{equation}
		\delta_{dLR}=-\frac{\omega}{2}\left(\mathrm{Im}[n_{L}]-\mathrm{Im}[n_{R}]\right)=-\frac{1}{2}\mathrm{Im}\left[V_{0}\right]=0, 
	\end{equation}
is null in this window, $\delta_{dLR} =0$, since the chiral parameter $V_{0}$ is considered real. Physically, it means that both RCP and LCP waves are absorbed at the same degree in the window $\omega<\omega_{\mathit{rad}}$, where the associated indices are complex. This behavior differs from the chiral magnetized plasma scenario, where the coefficient $\delta_{dLR}$ is non-null for different frequency ranges; see Ref. \cite{Filipe1}.
 
\begin{figure}
	\centering
	\includegraphics[scale=0.55]{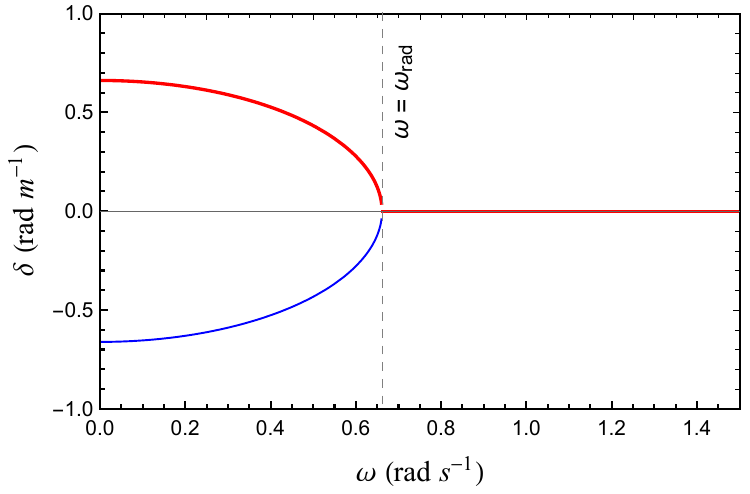}
	\caption{Red line: dichroism coefficient (\ref{dicro_TL_nER}), defined by the refractive indices $n_{E}$ and $n_{R}$ under the condition (\ref{condition1}). Blue line: dichroism coefficient (\ref{dicro_TL_nLM}), defined by the refractive indices $n_{L}$ and $n_{M}$ under the condition (\ref{condition1}). Here, we have used $V_{0}=(3/2)\omega_{p}$, and $\omega_{p}=1$~$\mathrm{rad}$~$s^{-1}$.}
	\label{DC_nEnR_1}
\end{figure}

For the indices $n_{E}$ and $n_{R}$, under the condition (\ref{condition1}), the circular dichroism coefficient is
 \begin{equation}
 \delta_{dER}=\begin{cases}
 \text{$ \omega \sqrt{R\left(\omega\right)}$}, &\quad\text{for $0<\omega<\omega_{rad}$},\\
 \text{$0$}, &\quad\text{for $\omega>\omega_{rad}$}.\\
 \end{cases}     \label{dicro_TL_nER}
 \end{equation}
Such a coefficient is non-null only for $\omega<\omega_{rad}$, being illustrated in Fig.~\ref{DC_nEnR_1}. It is also different from the one of a chiral magnetized plasma, which can be non-null for one or more intervals, as discussed in Ref.~\cite{Filipe1}.
 
Finally, we write the circular dichroism coefficient for the refractive indices $n_{L}$ and $n_{M}$, under the condition (\ref{condition1}),  
 \begin{equation}
\delta_{dLM}=\begin{cases}
\text{$ -\omega \sqrt{R\left(\omega\right)}$}, &\quad\text{for $0<\omega<\omega_{rad}$},\\
\text{$0$}, &\quad\text{for $\omega>\omega_{rad}$},\\
\end{cases}     \label{dicro_TL_nLM}
\end{equation}
which presents non-null and negative dichroism only for $\omega<\omega_{rad}$, as shown by the blue line in Fig.~\ref{DC_nEnR_1}, which corresponds to the mirrored behavior (opposite sign) of the coefficient (\ref{dicro_TL_nER}).

\section{Charge density oscillations \label{densityoscillations}}

Charge density oscillations can be analyzed from the set of equations for an unmagnetized plasma,
	\begin{gather}
		\frac{\partial n}{\partial t}+\mathbf{\nabla}\cdot\left(  n\mathbf{u}\right)
		=0,\label{plasmon}\\
		\frac{\partial\mathbf{u}}{\partial t}+\mathbf{u}\cdot\mathbf{\nabla u}%
		=\frac{q}{m}\mathbf{E} - \frac{1}{m n} \nabla P
		,\label{plasmons-3}%
	\end{gather}
	where $n$ is the electron number density, $\mathbf{u}$ is the electron fluid velocity field, $q$, and $m$ are the electron charge and mass, respectively. In addition, $P$ is the pressure, which describes the effects of the thermal kinetic energy of the electrons. For cold plasmas, one can neglect the pressure term in \eqref{plasmons-3} \cite{Bittencourt}. Considering small fluctuations around the electron density and electric field,
	\begin{subequations}
		\label{cold-plasma-fluctuations-from-equilibrium-1}
		\begin{align}
			n&=n_{0}+n',\\
			\mathbf{u}&=\mathbf{u}',\\ \mathbf{E}&=\mathbf{E}', \label{magneticp1}
		\end{align}
	\end{subequations} 
in Eqs. (\ref{plasmon}) and (\ref{plasmons-3}), the density perturbation turns out to be governed by the following differential equation:
\begin{align}
		\partial_{t}^{2} n' + \frac{e n_{0}}{m} \nabla \cdot {\bf{E}} - \frac{1}{m} 	\left( \frac{\partial P}{\partial n} \right)_{0} \nabla^{2} n'  &=0 . \label{plasmons-30}
\end{align}
Here, we have used $n'=\delta n$, ${\bf{u}}'=\delta {\bf{u}}$, and ${\bf{E}}'=\delta {\bf{E}}$ for the perturbations (physical quantities fluctutations). Assuming plane-wave ansatz in perturbative quantities and using the usual Gauss's law,  $(\nabla \cdot {\bf{E}})=en'/\epsilon_0$,equation  (\ref{plasmons-30}) yields
	\begin{equation}
		\omega^{2} = \omega_{p}^{2} + \frac{1}{m} \left( \frac{\partial P}{\partial n} \right)_{0} k^{2} , \label{plasmons-35}
	\end{equation}
which is the dispersion relation for the propagating plasma oscillations, known as Langmuir waves. In the absence of the pressure term in relation (\ref{plasmons-35}), it reduces to stationary oscillations at the plasma frequency, $\omega^{2}=\omega_{p}^{2}$, called Langmuir oscillations.
	Such oscillations are longitudinal to the electric field, since Eqs.~(\ref{plasmon}) and (\ref{plasmons-3}) provide the same direction for the charge velocity and the electric field \cite{Bittencourt}, that is,
	\begin{equation}
		\mathbf{u}'=\left[ \frac{i e}{m \omega} + \frac{i \epsilon_{0}k^2}{m n_{0} \omega e}  \left( \frac{\partial P}{\partial n} \right)_{0} \right]\mathbf{{E}}' . \label{uE}
\end{equation} 
Furthermore, the plasma oscillations derived from \eqref{plasmon}, up to first order in the linearized equations, yield the following relation for the charge density fluctuations:
	\begin{equation}
		n'= \frac{n_{0}}{\omega} ( {\bf{k}}\cdot {\bf{u}'}), \label{extra-continuity-1}
	\end{equation}
which, combined with (\ref{uE}), ensures that the density pertubations ($n'\ne0$) only occur for longitudinal waves, $\mathbf{k}\parallel\mathbf{{E}}'$. 

In this work, we consider a purely timelike vector $K_{AF}^{\mu}=(K_{AF}^{0}, {\bf{0}})$ in the MCFJ equations, which implies doing $\mathbf{k}_{AF}=0$ in Eqs. (\ref{Coulomb1}) and (\ref{Amp1}). In this context, we point out that Gauss's law remains unmodified, with the usual Langmuir dispersion relation (\ref{plasmons-35}) being preserved. In turn, Amperè's law yields modified dispersion relations for transverse RCP and LCP electromagnetic waves, with $\mathbf{k}\perp\mathbf{{E}}'$, as seen in Sec.~\ref{Propag.TL}. However,  there is no charge density perturbation $n'$ in this situation, in agreement with \eqref{extra-continuity-1}. 
Therefore, we conclude that the chiral factor $V_{0}$ does not affect the charge density fluctuations in this plasma system. We also point out that the circularly polarized propagating modes, LCP and RCP waves, do not alter the charge density fluctuations, which are longitudinal waves, as already explained.)

Additionally, one may infer how the chiral factor $V_0$ changes the equilibrium charge density of the plasma. This can be achieved by measuring, for a given frequency $\omega$, either $n_{L,E}$ or $n_{R, M}$, associated with the LCP and RCP propagating modes, respectively. Indeed, using Eqs.~(\ref{n-R-M-indices-1}) and (\ref{n-L-E-indices-1}), one finds
\begin{align}
n_{0} &= \frac{m \epsilon_{0}}{e^{2}} \omega^{2} \left(1-n_{l,r}^{2} \pm n_{l, r} \frac{V_{0}}{\omega} \right), \label{equilibrium-density-1}
\end{align}
where $n_{l} \equiv n_{L, E}$ and $n_{r} \equiv n_{R, M}$, in the frequency regime of real refractive indices. In the usual scenario, $V_{0}=0$, one obtains $n_{0} = (m \epsilon_{0} / e^{2}) \omega ^{2} (1 - n^{2})$ for the charge density in unmagnetized plasmas, with $n_{l}=n_{r}=n$.

From \eqref{equilibrium-density-1}, one has $n_{0} >0$ when $n_{l-} < n_{l} < n_{l+}$, where
\begin{align}
n_{l\pm} &= \frac{V_{0}}{2\omega} \pm \sqrt{ 1 + \frac{V_{0}^{2}}{4\omega^{2}}} . \label{extra-density-1}
\end{align}
Also, $n_{0} >0$ when $n_{r-} < n_{r} < n_{r+}$ with
\begin{align}
n_{r\pm} &= -\frac{V_{0}}{2\omega} \pm \sqrt{ 1 + \frac{V_{0}^{2}}{4\omega^{2}}} . \label{extra-density-2}
\end{align}
Thus, we observe that the regime where $\omega_{p}^{2} \ll \omega^{2}$ guarantees that $n_{l}$ and $n_{r}$ will be real.  Therefore, expression (\ref{equilibrium-density-1}) gives the charge density $n_{0}$ in terms of $n_{l}$ or $n_{r}$ for a given frequency $\omega >> \omega_{p}$.

\section{Final remarks \label{conclusion}}

In this paper, the propagation and absorption of the electromagnetic waves in unmagnetized plasma were analyzed in the case of the chiral axion electrodynamics stemming from the MCFJ timelike sector, which determines optical activity, negative refraction and RP sign reversion, a richer scenario than the one of a usual magnetized cold plasma. 

The Ampère's law of this theory includes the CME current term, $\mathbf{J}_{CME}=\left(K_{AF}\right)_{0}\mathbf{B}$, with the scalar background $\left(K_{AF}\right)_{0}$ representing the magnetic conductivity. The dispersion relation and corresponding refractive indices were derived by using the usual unmagnetized plasma permittivity and the modified Maxwell equations.

In Sec.~\ref{Propag.TL}, we obtained four modified refractive indices, $n_{R, M}$ and $n_{L, E}$, associated with RCP and LCP modes, respectively. The properties of such indices were analyzed and highlighted in the Secs.~\ref{secNR} - \ref{secNE} and Figs.~\ref{nRfig} - \ref{nEfig2}, where one can see that $n_{E}$ is the oppositive (negative) of $n_{R}$, and $n_{M}$ is the oppositive (negative) of $n_{L}$. Such indices manifest significant modifications in relation to the ones of a typical non-magnetized plasma.  For example, the refractive index $n_{R}$, illustracted in Fig.~\ref{nRfig}, possesses a negative refraction behavior in the range $0<\omega<\omega_{p}$, occurring a free negative propagation in the window $\omega _{\mathit{rad}}<\omega<\omega_{p}$. This feature is different from what takes place in the usual unmagnetized plasmas, where pure absorption happens in the interval $0<\omega<\omega_{p}$. Furthermore, in a magnetized chiral plasma scenario, negative refraction also emerges in the index associated with the RCP wave, as seen in Ref.~\cite{Filipe1}. The very low-frequency regime was analyzed in Sec.~\ref{section-helicons}, where RCP and LCP helicons were obtained, given by Eqs.~(\ref{helicons-15}) and (\ref{helicons-16}), respectively. In the present context, both RCP and LCP helicons can propagate simultaneously, a common point with the magnetized case \cite{Filipe1}.

Optical effects, including rotatory power and dichroism, were discussed in Sec.~\ref{secRP} under the conditions $V_{0}^{2}/4<\omega_{p}^{2}$ and $V_{0}^{2}/4>\omega_{p}^{2}$.  Despite yielding isotropic refractive indices, see Eqs. (\ref{n-R-M-indices-1}) and (\ref{n-L-E-indices-1}), the chiral factor induces birefringence (absent in usual unmagnetized plasmas), mimicking the optical role played by the magnetic field in cold plasmas. This birefringence is expressed in terms of the RP. For the indices $n_{R}$ and $n_{L}$, {the associated RP is constant, $\delta_{LR}=-V_{0}/2$, under any condition.} Such a value represents the high-frequency limit of the corresponding RP in a magnetized chiral plasma context, as shown in Eq.~(77) of Ref.~\cite{Filipe1}.  On the other hand, {a dispersive RP is defined for the indices $n_{R}$ and $n_{E}$, which is negative at low frequencies and positive at high frequencies, undergoing a sign reversion at the plasma frequency $\omega_{p}$, under both conditions, as depicted in Fig.~\ref{RP_nEnR}. The RP was also evaluated for the refractive indices $n_{L}$ and $n_{M}$, implying a negative coefficient, as illustrated in Fig.~\ref{RP_nLnM}.}

The effect of RP reversion has been reported in rotating plasmas \cite{Gueroult} as well, where the RP undergoes sign reversion and behaves with $1/\omega^2$ for high frequencies. RP reversion was also reported in the more involved scenario of an axion-like MCFJ magnetized chiral plasma, as shown in Figs.~15 and 16 of Ref.~\cite{Filipe1}. The axion term is thus identified as RP reversion inductor factor (in the absence or presence of the magnetic field). In this sense, the RP reversion may be a signature of chiral magnetic currents in pulsars, which can have relevance in astrophysics.

The dichroism effect was examined in Sec.~\ref{secDC},  with the coefficient $\delta_{dLR}$, associated with the indices $n_{L}$ and $n_{R}$, being null, which is in full compatibility with the constant RP (\ref{rptl}). For the indices $n_{E}$ and $n_{R}$, the coefficient $\delta_{dER}$ is non-null in the range $0<\omega<\omega_{rad}$ and null for $\omega>\omega_{rad}$, under the condition (\ref{condition1}). This behavior is very different from the magnetized case, where both coefficients,  $\delta_{dLR}$ and $\delta_{dER}$, are non-null for varied intervals. See Sec.~VI B in Ref. \cite{Filipe1}. 

 The plasma optical effects addressed in the present work could be measured in experimental setups involving the measurement of birefringence and rotatory power for the waves through the medium and frequency given in the propagation window.  For instance, the rotatory power given in Eqs. (\ref{rptl}), (\ref{rpeq2}), (\ref{rpeq2B}) can be probed by birefringence experimental devices, also constituting a route to determine the magnitude of the chiral factor, $V_0$. A sophisticated technique to measure such property was developed in Refs.~\cite{Birefringence1, Pajdzik2}. The system is based on a rotating polarizer, a quarter-wave plate, and an analyzer. The passing light is captured by a CDD camera, whereupon the measured data are evaluated by the computer. The intensity of the light detected depends on the $sin \, \alpha$,  with	$\alpha = 2\pi l (\Delta n)/\lambda$,
where $l$ is the sample thickness (plasma length traveled by the waves), and $\lambda$ is the vacuum wavelength of the incoming light. Thus, the device system \cite{Birefringence1, Pajdzik2} can be employed to determine the factor $\sin \, \alpha$ and measure the birefringence of traveling waves in the plasma of thickness $l$. This factor gives us the measured difference between the refractive indices of the material using $\Delta n = \alpha \lambda / (2 \pi l)$, which can be expressed in terms of the chiral factor $V_0$ and frequency by using the refractive indices (\ref{n-R-M-indices-1}) and (\ref{n-L-E-indices-1}), allowing us to determine the magnitude of the chiral factor. A similar technique was also used in Ref.~\cite{Geday}.

An interesting point is to establish a comparison among the present scenario, the usual magnetized plasma, and chiral MCFJ plasmas. Indeed, Table~\ref{tab:comparison-between-all-scenarios-2} compares four scenarios of cold plasmas: i) usual unmagnetized plasma, ii) usual magnetized plasma, iii) unmagnetized plasma ruled by the chiral MCFJ theory with $\mathbf{J}_{B}=K_{AF}^{0}\mathbf{B}$, iv) magnetized cold plasma ruled by the MCFJ theory with $\mathbf{J}_{B}=K_{AF}^{0}\mathbf{B}$, {with the magnetized cases defined for the Faraday configuration, $\mathbf{B}\parallel \mathbf{k}$. It points out relevant optical aspects, such as propagating modes, birefringence, RP reversion, absorption, and helicon modes. As well known, usual magnetized cold plasma presents birefringence and absorption, expressed in terms of RP and dichroism coefficient, but no RP reversion.} It differs from magnetized and unmagnetized cold plasma in MCFJ electrodynamics, where RP reversion occurs in both scenarios. Another difference is observed in propagating helicons, which occurs only for RCP waves in usual magnetized plasma, while both RCP and LCP can appear for magnetized and unmagnetized cold plasma governed by the MCFJ theory with $\mathbf{J}_{B}=K_{AF}^{0}\mathbf{B}$. For the Voigt configuration,  both magnetized cold plasma (usual and chiral MCFJ) exhibits elliptical modes, manifesting birefringence (phase shift) and absorption, as discussed in Ref. \cite{Filipe1}.

Finally, we have also examined the plasma charge density oscillations, ruled by Eqs. (\ref{plasmon}), (\ref{plasmons-3}) and the Gauss´s law. It was observed that MCFJ chiral factor $V_{0}$ does not modify the charge density fluctuations but can alter the equilibrium charge density, $n_0$, as shown in \eqref{equilibrium-density-1}.


\begin{widetext}


	\begin{table}[H] 
	\caption{Propagation properties of four distinct scenarios of cold plasmas. The symbol ``$...$" means that the entry does not apply to the mentioned feature and plasma model. The magnetized scenarios are for the Faraday configuration, $\mathbf{B}\parallel \mathbf{k}$.}
		\centering
			\begin{tabular}{ C{1.9cm}  C{3cm} C{3.8cm} C{4.2cm}  C{4.3cm} C{.5cm} }
				\toprule \\[0.01ex]
				\\[-4.5ex]
				& \textbf{Unmagnetized plasma in usual electrodynamics} 	& \textbf{Magnetized plasma in usual electrodynamics} & \textbf{Unmagnetized {chiral MCFJ plasma} with $\mathbf{J}_{B}=k_{AF}^{0}\mathbf{B}$} & \textbf{Magnetized  {chiral MCFJ plasma} with $\mathbf{J}_{B}=k_{AF}^{0}\mathbf{B}$}   \\[0.6ex]
				\colrule \\[0.6ex]
				\\[-4.5ex]
				Propagating modes  & linear mode & RCP  for $n_{-}$, LCP for $n_{+}$ & RCP  for $n_{R, M}$, LCP for $n_{L, E}$  & RCP  for $n_{R, M}$, LCP for $n_{L, E}$, see Ref. \cite{Filipe1}
				\\[0.6ex]
				\colrule \\[0.1ex]
				\\[-4ex]
				Birefringence & ...   &  RP     & RP (\ref{rptl}) and (\ref{rpeq2})  & RP $\delta_{LR}$ and $\delta_{ER}$, see Ref. \cite{Filipe1}
				\\[0.6ex]
				\colrule \\[0.1ex]
				\\[-4.5ex]
				RP {reversion} &    ...  & no &  yes  & yes  \\ [0.6ex]
				\colrule \\[0.1ex]
				\\[-4ex]
				Absorption & yes (conventional)  & yes (dichroism) &   yes, \, $ \delta_{dER}$ (\ref{dicro_TL_nER}) & yes (dichroism)  \\ [0.6ex]
				\colrule \\[0.1ex]
				\\[-4ex]
				Helicons & ... & RCP & RCP and LCP, enabled by the chiral factor  & RCP and LCP, see Ref. \cite{Filipe1} \\ [0.6ex]
				\botrule
			\end{tabular}
		\label{tab:comparison-between-all-scenarios-2}
	\end{table}
\end{widetext}

	\begin{acknowledgments}
	
The authors express their gratitude to FAPEMA, CNPq, and CAPES (Brazilian research agencies) for their invaluable financial support. M.M.F. is supported by FAPEMA APP-12151/22, CNPq/Produtividade 317048/2023-6 and CNPq/Universal/422527/2021-1. P.D.S.S. is grateful to grant CNPq/PDJ 150584/23 and FAPEMA APP-12151/22. Furthermore, we are indebted to CAPES/Finance Code 001 and FAPEMA/POS-GRAD-02575/21.

	\end{acknowledgments}

\end{document}